\documentclass[conference]{IEEEtran}

\usepackage{eso-pic}
\AddToShipoutPictureBG*{%
  \AtPageUpperLeft{%
    \raisebox{-1cm}{%
      \makebox[\paperwidth]{\scriptsize Accepted at 2026 IEEE 31st International Conference on Emerging Technologies and Factory Automation (ETFA)}}}}
      
\AddToShipoutPictureBG*{%
  \AtPageLowerLeft{%
    \raisebox{0.5cm}{%
      \makebox[\paperwidth]{%
        \parbox{0.85\paperwidth}{\scriptsize\centering
          \copyright\ 2026 IEEE. Personal use of this material is permitted. Permission from IEEE must be obtained for all other uses, in any current or future media, including reprinting/republishing this material for advertising or promotional purposes, creating new collective works, for resale or redistribution to servers or lists, or reuse of any copyrighted component of this work in other works. DOI: to appear
        }}}}}

\usepackage{cite}
\usepackage{amsmath,amssymb,amsfonts}
\usepackage{algorithmic}
\usepackage{graphicx}
\usepackage{xcolor}
\usepackage{booktabs}
\usepackage{multirow}
\usepackage{siunitx}

\usepackage{pgfplots} 
\pgfplotsset{compat=1.18}
\definecolor{colPlate}{HTML}{64b4ff}
\definecolor{colAssembly}{HTML}{ffaa3c}
\definecolor{colRotation}{HTML}{ff78c8}
\definecolor{colQuality}{HTML}{78e678}
\definecolor{colRecycling}{HTML}{dc5050}
\definecolor{colAGV}{HTML}{b48cff}
\definecolor{colBackbone}{HTML}{8c8ca0}

\def\BibTeX{{\rm B\kern-.05em{\sc i\kern-.025em b}\kern-.08em
    T\kern-.1667em\lower.7ex\hbox{E}\kern-.125emX}}
\begin{document}

\title{LLM-Driven Multi-Agent Control for Skill-Based Smart Manufacturing}

\author{  
  \IEEEauthorblockN{
  Kay Köhle\IEEEauthorrefmark{1}\IEEEauthorrefmark{2}, 
  Darko Anicic\IEEEauthorrefmark{2}, 
  Thomas A. Runkler\IEEEauthorrefmark{1}\IEEEauthorrefmark{2},
  René Graf\IEEEauthorrefmark{3}}
  \IEEEauthorblockA{\IEEEauthorrefmark{1}Technical University of Munich, Munich, Germany} 
  \IEEEauthorblockA{\IEEEauthorrefmark{2}Siemens AG, Foundational Technologies, Munich/Garching, Germany}  
  \IEEEauthorblockA{\IEEEauthorrefmark{3}Siemens AG, Digital Industries, Fürth, Germany}
  \IEEEauthorblockA{Email: \texttt{kay.koehle@siemens.com}}
}  

\maketitle

\begin{abstract} 

Factories are shifting toward smaller lot sizes with high product customization, requiring frequent re-programming of flexible and reconfigurable automation systems.
LLM-based agents can be deployed in two complementary roles:
offline, they generate deterministic production sequences, reducing programming effort; online, they operate live machines and handle unforeseen runtime faults that static programs cannot anticipate.
We propose a solution in which each factory module is paired with a dedicated LLM-based agent and an MCP tool server that exposes the module's skills via OPC UA method calls, with agents coordinating over MQTT and grounded by real time updates of the factory state.
We compare three agent architectures (orchestrator, peer-to-peer, and monolithic) across nine production challenges of increasing complexity in a simulation of a physical six-module hexagonal factory, including silent hardware fault detection.
The monolithic and peer-to-peer architectures both achieve the highest mean solve rate (93\%), while the orchestrator uniquely resolves a silent conveyor-belt fault in all ten runs by autonomously rerouting plates around the blocked segment.
All architectures exhibit emergent fault-diagnosis behavior without any explicit failure-handling logic, establishing standardized MCP tooling, MQTT-based inter-agent communication, and real-time state injection as a viable and reproducible foundation for LLM-programmed smart manufacturing.
\end{abstract}

\begin{IEEEkeywords}
Multi-Agent System, Large Language Model, OPC UA, MQTT, Model Context Protocol, Industrial Automation, Smart Manufacturing, Cyber-Physical System
\end{IEEEkeywords}

\section{Introduction} 
Industrial automation systems are undergoing a profound transformation driven by the shift toward high-mix, low-volume manufacturing, which requires rapidly reprogrammable production systems.
LLM-based agents can generate such production sequences, reducing manual programming effort.
Additionally, while classical deterministic schedulers handle fixed production sequences reliably, they require every recoverable condition to be anticipated and hard-coded.
Silent hardware faults, where a method reports success, but produces no physical effect, fall outside this model entirely, as they cannot be enumerated in advance.
Here, LLM agents that directly control the modules inside a production system can reason over discrepancies between expected and actual factory state, attributing failures and rerouting production without any explicit pre-programmed failure-handling logic.
Early work has demonstrated that LLM-based agents can interpret natural-language task descriptions and interact with cyber-physical production systems, with Xia et al.~\cite{Xia} showing how agents can coordinate industrial modules through event-driven reasoning over OPC UA-based digital twins.
Hofmann et al.~\cite{LLMandOPCUA} further show that LLMs can control real machines by invoking OPC UA methods as callable tools, while Xu et al.~\cite{LLMasController} demonstrate LLM-based control in robotic feedback loops, establishing LLMs as credible general-purpose control components.
In parallel, the Model Context Protocol (MCP) has introduced a standardized way to expose machine capabilities to LLMs.
Vieira Da Silva et al.~\cite{MCPinManufacturing} show how MCP can formalize these capabilities within manufacturing settings, enabling LLMs to interact with machines and skills through semantically enriched, self-describing tool interfaces rather than unstructured natural-language heuristics.
At the communication layer, SUPCON Global has demonstrated LLM agents operating a simulated factory over MQTT, arguing that JSON-structured publish/subscribe messaging is well-suited for scalable multi-agent coordination~\cite{supcon2025factoryagent, supcon2025mqtt}.
Despite these developments, significant gaps remain.
Existing approaches either adopt monolithic agent designs without standardized tool interfaces, use custom event-driven prompting without a structured inter-agent communication protocol, or demonstrate MQTT-based coordination without connecting it to OPC UA as a machine-level actuation layer.
No prior work combines all three: structured agent architectures, MCP-based modular tooling, and OPC UA as the standardized execution interface.
This paper addresses these gaps by comparing agent architectures for LLM-driven industrial automation in which each factory module is paired with a dedicated MCP tool server that exposes its skills via OPC UA method calls, and agents coordinate over MQTT.
We evaluate three architectures (orchestrator, peer-to-peer, and monolithic) across nine production challenges of increasing complexity in a Python-based simulation of a factory with six modules.
In the challenges, the monolithic and peer-to-peer architectures achieve the highest mean solve rate of 93\%, while the orchestrator uniquely solves the transport-fault challenge at 100\% by leveraging its global view of the factory.

\section{Background}
\label{sec:bg}

\subsection{Skill-Based Manufacturing and OPC UA}
Skill-based manufacturing organizes machine functionality into discrete, callable skills that abstract away internal implementation complexity~\cite{skills}.
OPC UA is a platform-independent communication standard for industrial automation that combines a client-server interaction model with a structured information model~\cite{opcua}.
An OPC UA server exposes device data and functionality as a typed, hierarchical address space of objects, variables, and methods, with methods specifically serving as the standardized, invocable interface for a module's skills.


\subsection{LLM-based Agents and Model Context Protocol (MCP) Tools}

An LLM-based agent is a system in which a large language model operates in a perception–reasoning–action loop~\cite{yao2023react}, meaning it receives context from its environment and selects tool actions and iterates until a goal is reached.
This makes LLM-based agents suitable for tasks that require decision-making and interaction with external systems.
MCP is an open standard that introduces a client-server architecture for tool use, providing LLMs with a structured interface to invoke external functionality~\cite{mcp2024}, such as aforementioned OPC UA methods.



\section{Related Work}
\label{sec:related}

Research on industrial automation increasingly explores the use of LLMs for planning, decision-making, and execution.
The expectation is that the LLMs' reasoning capabilities enable them to flexibly handle unforeseen faults in modular production systems, a challenge that has long been recognized in resilient and agent-based manufacturing research~\cite{Pulikottil2023AgentManufacturing}.
An early demonstration of LLM-enabled multi-agent manufacturing was presented by Lim et al.~\cite{Lim2024}, who integrated GPT-4 into an orchestrator architecture to allocate G-code operations across simulated CNC machines via natural language function calls.
Zhao et al.~\cite{Zhao2026} deploy a multi-agent LLM system on a physical shopfloor, in which each machine is controlled by an agent composed of specialized modules that negotiate machine selection in natural language.
The most directly relevant work is Xia et al.~\cite{Xia}, who propose a multi‑agent architecture for planning and executing tasks in a modular production system using digital twins and OPC UA.
Vieira Da Silva et al.~\cite{MCPinManufacturing} use the Model Context Protocol (MCP) to expose manufacturing skills as semantically enriched LLM-callable tool descriptions, focusing on semantic interoperability and capability modeling rather than runtime agent coordination.

Across these works, existing approaches either rely on proprietary cloud APIs and unstructured prompt-based interfaces~\cite{Lim2024, Zhao2026, Xia}, evaluate only single-step action prediction without cross-module fault routing~\cite{Xia}, or address capability modeling without runtime coordination~\cite{MCPinManufacturing}.
In contrast, our work combines a multi-agent architecture, MCP-based per-agent tool servers, OPC UA as the standardized execution interface, and MQTT as the inter-agent communication protocol.

\section{Concept}
\label{sec:concept}

\begin{figure}[ht]
\centerline{\includegraphics[keepaspectratio]{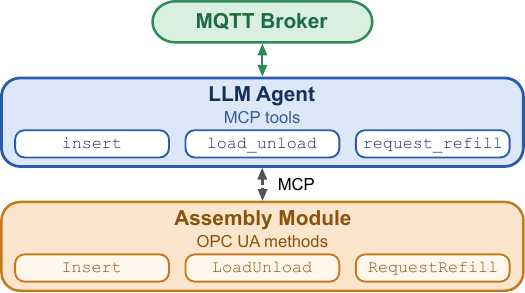}}
\caption{For each factory module, OPC UA method calls expose the module's skills, which can be invoked by the module's corresponding LLM-agent over MCP tools. The agents communicate over MQTT.}
\label{fig:workflow}
\end{figure}

This section presents our architecture for LLM-agent controlled modular production systems.
We describe how factory module skills are exposed as agent-callable tools, how multiple agents coordinate over a publish/subscribe communication layer, and how the current physical state of the factory is externally injected to ground the agents' reasoning.

\subsection{Factory Module Agent Architecture}

Figure~\ref{fig:workflow} visualizes the architecture of a single agent.
Each agent is a self-contained unit paired with a factory module, with its OPC UA methods exposed to the LLM as tools, and agents coordinating over an MQTT broker.

In our prototype, each factory module is assigned a dedicated MCP server that wraps the module's OPC UA methods as MCP tools.

MCP was chosen for its native integration with major LLM inference frameworks and its growing ecosystem.
Beyond this practical motivation, MCP naturally enforces the separation of concerns central to skill-based manufacturing.
The MCP server is the only component that knows the OPC UA endpoint address and method signatures, while the LLM agent knows only the tool's name, description, and schema.
As a result, replacing or extending a module's functionality requires only updating its MCP server, with no changes to the agent orchestration logic.
Although the implementation uses MCP, the underlying architectural principle of using one tool server per module, with typed schemas is protocol-agnostic and applicable to any production system with OPC UA-accessible module skills.

\begin{figure*}[!htbp]
\centerline{\includegraphics[keepaspectratio]{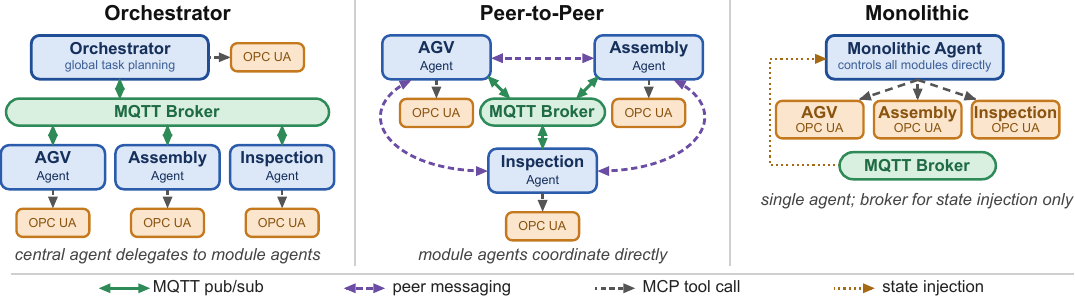}}
\caption{Overview of the three agent architectures.}
\label{fig:architectures}
\end{figure*}

\subsection{Multi-Agent Coordination via MQTT}
We arrange agents in three distinct architectures, illustrated in Figure~\ref{fig:architectures}.
In the \textit{orchestrator} architecture, a central agent holds a global view of the factory and delegates tasks to dedicated module agents.
In the \textit{peer-to-peer} architecture, module agents communicate directly with one another without a central coordinator.
In the \textit{monolithic} architecture, a single agent controls all modules directly.
These three patterns expose different trade-offs in coordination overhead, fault resilience, and reasoning scope, motivating a systematic comparison.
The detailed implementation of each architecture is described in Section~\ref{sec:implementation}.


MQTT is an established industrial standard for machine-to-machine communication, widely deployed on resource-constrained edge devices\cite{IoTProtocolSurvey}.
Unlike OPC UA's traditional client-server model, which requires agents to maintain direct endpoint references, MQTT's publish/subscribe mechanism naturally maps to a multi-agent topology, because agents are decoupled via a broker, new agents can join without modifying existing ones, and topic-based routing natively supports one-to-many broadcast semantics.
While OPC UA's Pub/Sub extension offers a hybrid alternative by combining OPC UA's typed information model with MQTT transport semantics, in our prototype we preferred plain MQTT for its simplicity and broader native support in edge environments.

\subsection{Real-Time State Injection}
\label{subsec:tracker}
We track the effects of OPC UA method calls on the system in an internal state tracker.
The state tracker maintains a live snapshot of plate locations, belt occupancy, block inventory, and inspection results, acting as an event-driven aggregator over the stream of skill results.
Each skill completion event, received via MQTT, carries the information needed to update the state, for example \texttt{MovePlate} updates the plate's location, while \texttt{LoadUnload} updates the occupancy of the corresponding module.
At each inference step, this state is passed to the LLM agents by prepending it to the input prompt as a structured context block.
This relieves LLMs from having to reconstruct the state themselves from the conversation history, with the added benefit of enabling aggressive token-conserving context-history trimming.

The internal state also drives the dynamic tool constraint mechanism.
A tool availability manager evaluates the current simulation state and publishes the set of physically valid tools to each agent over MQTT.
The agent's tool list is filtered before each LLM call, preventing the model from attempting operations that would be rejected by the underlying controllers.
For example, an inspection cannot be invoked before the plate has been loaded into the module.

The mechanisms of standardized tool abstraction, publish/subscribe communication, and externally injected state together allow the agents to reason and act locally while the system as a whole executes coherent multi-step production workflows.

\section{Implementation}
\label{sec:implementation}
Our prototype consists of two main parts:
A simulated factory and an agent orchestration system that controls it.

\subsection{Factory Simulation}


The prototype's simulation environment is modeled after a small real-world industrial demonstration lab equipped with a physical manufacturing line.
The manufacturing lab consists of seven modules, including a hexagonal conveyor belt as the backbone.
The conveyor consists of an inner belt for inter-module transit and at each edge it has an outer belt that serves as the docking interface for the six peripheral modules.
Plates circulate on the inner belt and are transferred to the outer belt when a module needs to interact with them.
Modules can only load and unload plates from the outer belt, meaning a plate must first be positioned at the correct outer belt segment before any module operation can be invoked.
The lab manufactures a product that consists of five wooden blocks on a plate.

\begin{figure}[ht]
\centerline{\includegraphics[width=0.2\textwidth]{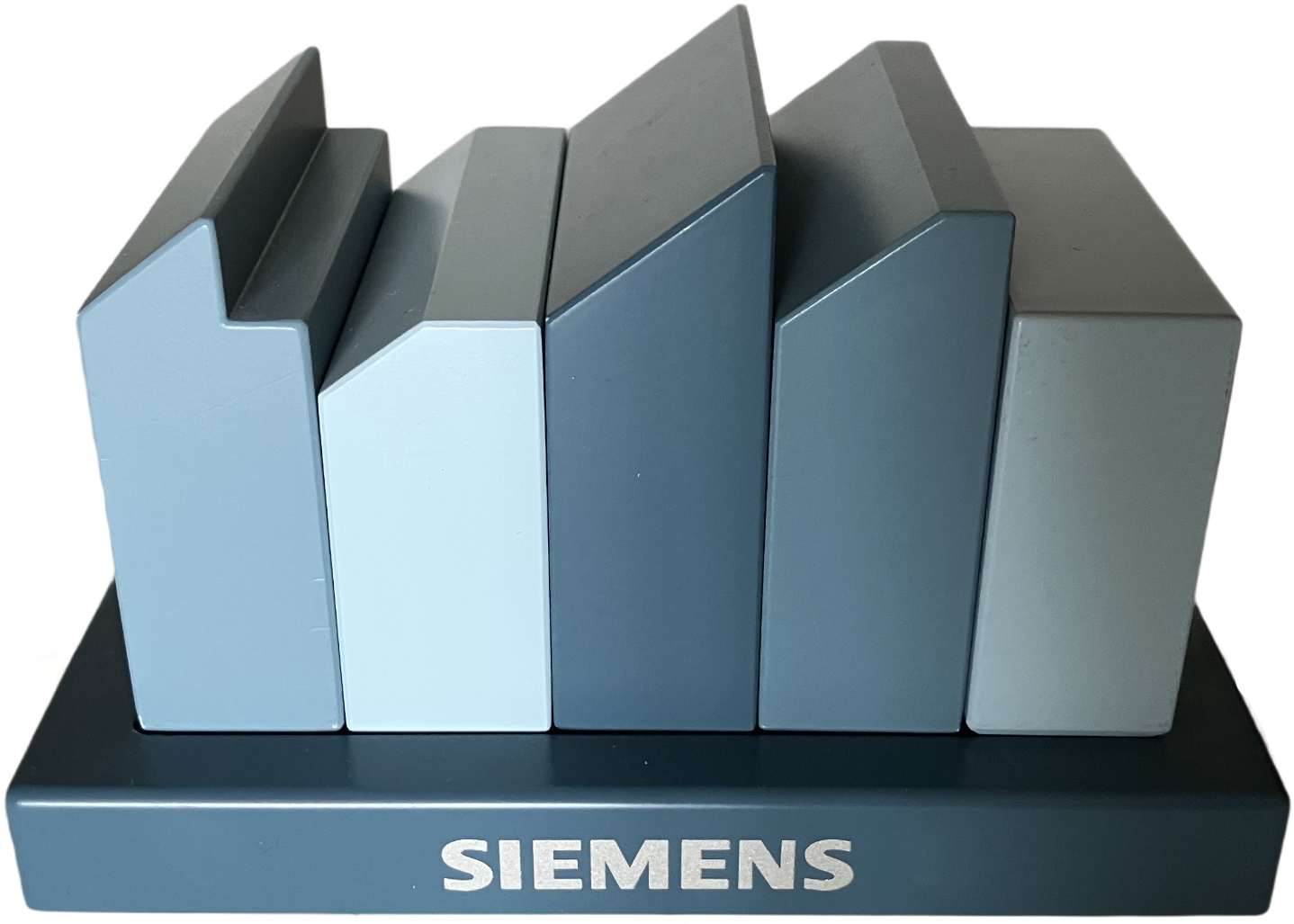}}
\caption{Example product featuring five wooden blocks of different shapes on a plate, placed at \ang{0} or \ang{180} rotation.}
\label{fig:product}
\end{figure}

A typical order specifies a plate with up to five wooden blocks, each of ten different types placeable at \ang{0} or \ang{180} rotation. The Plate Dispenser Module places an empty plate on the belt, which is routed through the relevant Assembly, Inspection, and Recycling or Delivery Modules in sequence.
If a product is found faulty at inspection, blocks are removed at the Recycling Module and the assembly process restarts.
Each module has its own OPC UA server, which exposes the module's skills as OPC UA methods (see Table \ref{tab:module-tools}).
The hexagonal backbone module exposes a single method, \texttt{MovePlate(edge, direction, source, destination, plateID)}, which moves a plate one step along the belt.
The \texttt{edge} parameter identifies one of the six sides of the hexagon, and \texttt{direction} specifies the rotational direction of travel.
The \texttt{source} and \texttt{destination} parameters specify whether the plate is on the inner or outer belt at each end of the move, allowing the method to express both transit steps along the inner belt and transfers between the inner and outer belt at a given edge.
The Plate Dispenser Module can place or remove an empty plate onto or from its outer belt segment via \texttt{PlacePlate()} and \texttt{RemovePlate()}.
All other peripheral modules expose a \texttt{LoadUnload(source, destination)} method that fetches a plate from its corresponding outer belt segment into the module or returns it afterward.
Only once a plate has been loaded into a module can that module's special-purpose operation be invoked, such as \texttt{Insert()} for assembly, \texttt{Inspect()} for quality control, \texttt{Recycle()} for block removal, or \texttt{Deliver()} for AGV handoff.

\begin{table}[ht]
\centering
\caption{Factory modules and their corresponding OPC UA methods.}
\label{tab:module-tools}
\footnotesize
\begin{tabular}{p{1.5cm} p{5.5cm}}
\toprule
\textbf{Module} & \textbf{OPC UA Method} \\
\midrule
Hexagonal Backbone
    & \texttt{MovePlate(edge, direction,} \newline
      \texttt{source, destination, plateID)} \\
\midrule
\multirow{2}{*}{\parbox{1.5cm}{Plate \\ Module}}
    & \texttt{PlacePlate()} \\
  & \texttt{RemovePlate()} \\
\midrule
\multirow{2}{*}{\parbox{1.5cm}{Assembly \\ Module}}
    & \texttt{LoadUnload(source, destination)} \\
  & \texttt{Insert(magazineId, location)} \\
  & \texttt{RequestRefill(magazineId)} \\
\midrule
\multirow{2}{*}{\parbox{1.5cm}{Assembly \\ Module (Rot.)}}
    & \texttt{LoadUnload(source, destination)} \\
  & \texttt{Insert(magazineId, location,} \newline
    \texttt{rotation)} \\
  & \texttt{RequestRefill(magazineId)} \\
\midrule
\multirow{2}{*}{\parbox{1.5cm}{Quality \\ Inspection}}
    & \texttt{LoadUnload(source, destination)} \\
  & \texttt{Inspect()} \\
\midrule
\multirow{2}{*}{\parbox{1.5cm}{Recycling \\ Module}}
    & \texttt{LoadUnload(source, destination)} \\
  & \texttt{Recycle()} \\
\midrule
\multirow{2}{*}{\parbox{1.5cm}{AGV \\ Delivery}}
    & \texttt{LoadUnload(source, destination)} \\
  & \texttt{Deliver(plateId)} \\
\bottomrule
\end{tabular}
\end{table}

The simulation is implemented in Python and exposes mock OPC UA servers that manipulate the simulated factory's state in response to method calls.
After each state change, the simulation publishes a structured event over MQTT on the \texttt{factory/sim/\{event\_type\}} topic, which updates the internal state tracker.
As detailed in Section~\ref{subsec:tracker}, the factory state is prepended as context to every LLM inference turn, giving each agent an up-to-date view of the physical world without requiring the LLM to maintain state across turns.
We investigate the benefit of the internal state tracker in a small ablation study.

\begin{figure}[ht]
\centerline{\includegraphics[width=0.49\textwidth]{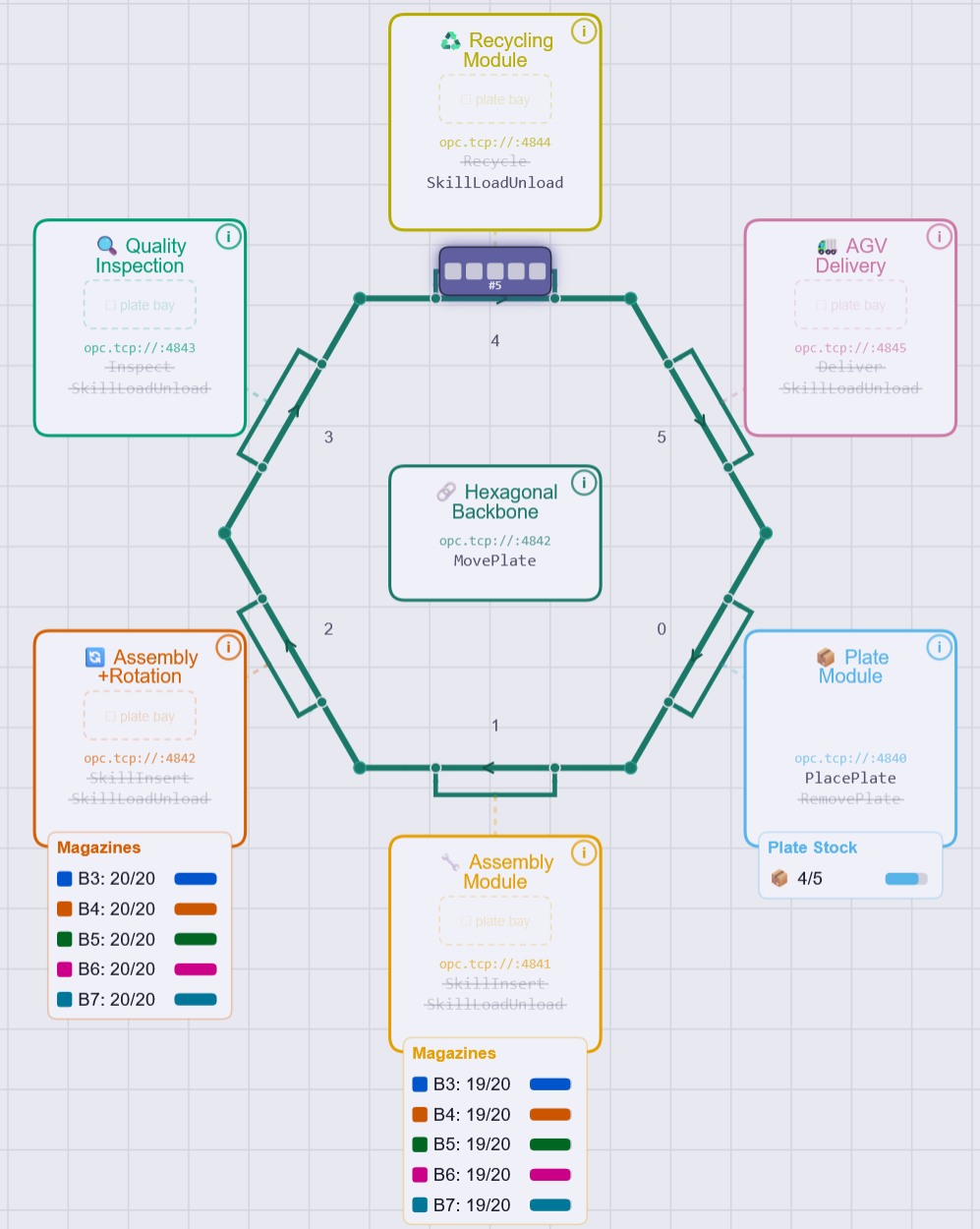}}
\caption{Real-time dashboard of the factory simulation. The six outside modules dock onto the hexagonal conveyor belt backbone in the middle. The ToolAvailabilityManager dynamically enables each module's operation tools based on the factory's state. As shown in the screenshot, the only permitted operations are loading the plate into the recycling module, moving the plate, or placing an additional plate onto the conveyor belt.}
\label{fig:dashboard}
\end{figure}

A real-time web dashboard visualizes the factory state by subscribing to the same MQTT simulation events, providing a live 2D view of plate positions and module activity during experiments (see Fig.~\ref{fig:dashboard}).
The web dashboard and simulation layer are intentionally decoupled from the agent architecture.
Any component that exposes an OPC UA interface can serve as a drop-in replacement for the mock servers, including real physical hardware or higher-fidelity digital twins, as demonstrated in \cite{VirtuallyCommissionedSemanticRecipes}.

\subsection{Agent Implementation}

This section details the concrete implementation of the three architectures introduced in Section~\ref{sec:concept}.
The first architecture assigns an agent to each module, with a dedicated orchestrator agent coordinating all module agents.
In our single-backbone prototype, this orchestrator also controls the conveyor belt, since it is the central transport medium.
We additionally implemented a peer-to-peer architecture and a monolithic architecture for systematic comparison.
In the peer-to-peer architecture, the modules can control the conveyor belt at the edge they are docked to, to send plates to their adjacent modules.
In the monolithic architecture, a single agent has all the tools to control all modules of the factory, much like an LLM-driven MES.

\subsubsection{Orchestrator Architecture}

The backbone agent serves as the orchestrator, coordinating the six specialist agents.
The orchestrator's system prompt encodes the full routing logic, specifying which module at which edge handles which block type, the load/unload protocol, and the decision rules for routing after inspection.
The current factory state, including plate locations, belt occupancy, assembly progress, and inspection results, is prepended to every inference turn to provide the orchestrator with an up-to-date view of the physical world.
The prompt for specialist agents encodes the module's skills, its edge number, its magazine contents, and block types.
The orchestrator exchanges messages with the specialist agents via a local MQTT broker, using a dedicated command topic and a dedicated reply topic per specialist.
When the orchestrator wishes to delegate a task, it publishes a message to the specialist's command topic.
On receiving a message from the orchestrator, the specialist agent calls the LLM with the current tool list, executes any tool calls via MCP, feeds results back to the LLM, and repeats until the LLM produces a plain text response, which is published back over MQTT.
Once the specialist has completed its task, it publishes its response to its own reply topic, closing the request–response cycle.
The backbone subscribes to all specialist reply topics.
Each agent has its dedicated MCP tool server, which exposes the module's OPC UA methods as tools.
To prevent agents from attempting physically impossible operations (e.g., loading a non-existent plate), a ToolAvailabilityManager continuously evaluates the simulation state and publishes the set of currently valid tools, preventing such operations without requiring the LLM to reason about preconditions.
The ToolAvailabilityManager is also used in the other two architectures, with adaptations for each.

\subsubsection{Peer-to-Peer Architecture}

In the peer-to-peer architecture, only the six outside module agents are deployed, without the central conveyor belt agent as an orchestrator.
Each agent is identical in structure to the backbone specialist agents and is paired with the same dedicated MCP server, exposing only its own module's OPC UA methods.
In addition to the MCP tools, each agent is augmented with a \texttt{send\_message} tool, which allows the LLM to route a message to a peer agent by name.
This tool publishes the payload to the target agent's dedicated MQTT topic. Incoming peer messages are injected as user-role turns with a \texttt{[From sender]} prefix, enriched with the current factory state, and then processed through the standard tool-calling loop.
The agent's response may contain further \texttt{send\_message} calls that are dispatched during the same loop, establishing a chain of bilateral handoffs rather than a centrally orchestrated sequence.
A per-agent lock serializes incoming messages to prevent corruption of the conversation history when multiple peers send messages simultaneously.
Each agent's system prompt encodes its edge number, its module's tool signatures, its constrained peer graph (i.e.\ only the subset of agents it is permitted to address), the relevant slice of the production orders, and the coordination protocol.
After completing a processing step, the agent moves the plate to the next edge and notifies the downstream peer.
The plate module agent is designated as the initiator and starts production autonomously after connecting.

\subsubsection{Monolithic Architecture}

The monolithic architecture uses a single agent with direct access to every factory module.
Its MCP server aggregates all module tools into a single endpoint, disambiguating same-named operations across modules with a module-name prefix.
The agent runs the same tool-calling loop as the specialist agents:
It calls the LLM with the full tool list, executes any tool calls via MCP, feeds results back, and repeats until the LLM produces a plain-text response.
The system prompt encodes the complete factory layout (all six edges and their roles), the full set of module tools with their signatures, the step-by-step workflow for processing a plate from placement through assembly, inspection, and delivery or recycling, and all production orders.
The factory state is prepended to every inference turn in the same format as in the backbone architecture.
Because the agent must reason across all modules in a single context, the history window and the maximum tool-call budget per turn are significantly larger than in the multi-agent architectures.
To keep the context window from overflowing, outdated factory states are removed from the history. 

\section{Experiments}
\label{sec:experiments}

We evaluate the multi-agent factory controller across nine increasingly complex challenges.
Each production challenge was run 10 times and the means are reported across successful runs.
All experiments were conducted on the Python-based factory simulation described in Section~\ref{sec:implementation}.
Gemma4:31B was selected as the LLM for its native support for function calling, strong agentic capabilities, and open weights under the Apache 2.0 license.
Inference was performed locally via Ollama using the 30.7B-parameter dense model with Q4\_K\_M (20 GB) quantization on an Nvidia RTX 6000 Pro Blackwell Workstation GPU rather than through a cloud provider, as datacenter load variability and network latency would introduce uncontrollable timing factors that could confound experimental results.
Temperature was set to 0.3 rather than 0 to allow agents to explore different recovery strategies across repeated runs of the silent-fault challenges.

\subsection{Production Challenges}
\label{subsec:challenges}

The challenges progress from basic pipeline verification (1–6) to hardware-level silent faults requiring autonomous diagnosis (7–9), each isolating or combining a distinct system capability.

\textbf{1) Single-Type Order, Single Module (Baseline).}
Five identical blocks (type~5).
Verifies the minimum viable pipeline: 
plate placement, routing to a single assembly station, five insertions, inspection, and AGV delivery.

\textbf{2) Multi-Type Order, Single Module.}
One block of each of five different types, all in one assembly module.
The agent must map block types to magazine IDs and complete assembly in a single module station visit.

\textbf{3) Multi-Type Order, Two Modules, With Rotation.}
Three block types from the first assembly module, two from the assembly with rotation module (one block at \ang{180}).
Requires cross-module routing and correct orientation handling.

\textbf{4) Multi-Type Order, Two Modules, With Refill.}
Same as Challenge 3, but magazine~1 of the assembly with rotation module starts empty.
The agent must call \texttt{request\_refill(magazine\_id=1)}, await operator confirmation, then resume.

\textbf{5) Multi-Type Order, Two Modules, With Recycling.}
Same order as Challenge~3, but the first assembly attempt contains a defective block.
Quality inspection fails and the plate is recycled; the second attempt succeeds.
This experiment exercises the full \textit{inspect}~$\rightarrow$~\textit{recycle}~$\rightarrow$~\textit{reassemble} loop.

\textbf{6) Two Parallel Multi-Type Orders, Two Modules, With Rotation.}
Two orders are active simultaneously, with order identity tracked implicitly via plate IDs.
The first exclusively requires blocks from the first assembly module, while the second requires blocks from the second assembly with rotation module (two at \ang{180}).
The agents must interleave transport across the modules and track per-plate order assignments.

\textbf{7) All Edge-1 Blocks Defective, Autonomous Fallback.}
Same order as Challenge~2, but the first assembly module contains only defective blocks, while the assembly with rotation module carries the same types with good blocks.
The agents must infer the pattern of inspection failures and reroute assembly without any explicit fault signal.

\textbf{8) Defective Assembly Module (Silent Failure).}
Same order as Challenge~2, but the first assembly module's \texttt{skill\_insert} reports success even though it places no blocks.
The agents must cross-verify via the quality inspector, attribute the failure to the assembly module, and coordinate rerouting to the second assembly module.

\textbf{9) Defective Conveyor Belt (Silent Failure).}
A silent fault is injected on the clockwise belt segment from edge~0 to edge~1.
\texttt{move\_plate} reports success but the plate stays at edge~0.
The agents must infer the transport failure and reroute counter-clockwise.

\subsection{Evaluation Metrics}
Each challenge is evaluated across the following metrics:
\begin{enumerate}
    \item \textbf{Solve Rate (SR)} — percentage of runs in which the agents successfully fulfilled the production order.
    \item \textbf{Time} — wall-clock time from the first agent action to successful product delivery.
    \item \textbf{Tokens} — total input and output tokens consumed per run, broken down per agent, as a measure of operational cost and context efficiency.
    \item \textbf{Tool Calls} — Number of tool calls used to fulfill the production order.
\end{enumerate}

\section{Results and Discussion}
\label{sec:results}

\begin{table*}[htbp]
\centering
\caption{Evaluation results of the agent architectures across 9 production challenges (mean over 10 runs). \\Opt = optimal tool calls; Calls = Tool calls on solved runs (min--max across runs).\\Bottom row: mean $\pm$ std across the nine challenge means. Tool call ranges are omitted from the summary row as min-–max ranges are not comparable across challenges with different Opt values.}
\label{tab:results-combined}
\sisetup{separate-uncertainty=true, mode=text, detect-weight=true}
\begin{tabular}{c c S[table-format=3] c S[table-format=3(2)] S[table-format=3(2)] S[table-format=3] c S[table-format=3(2)] S[table-format=3(2)] S[table-format=3] c S[table-format=3(2)] S[table-format=3(2)]}
\toprule
\textbf{\#} & {\textbf{Opt}} & \multicolumn{4}{c}{\textbf{Orchestrator}} & \multicolumn{4}{c}{\textbf{Peer-to-Peer}} & \multicolumn{4}{c}{\textbf{Monolithic}} \\
\cmidrule(lr){3-6} \cmidrule(lr){7-10} \cmidrule(lr){11-14}
 &  & {\textbf{SR (\%)}} & \textbf{Calls} & {\textbf{Time (s)}} & {\textbf{Tok. (k)}} & {\textbf{SR (\%)}} & \textbf{Calls} & {\textbf{Time (s)}} & {\textbf{Tok. (k)}} & {\textbf{SR (\%)}} & \textbf{Calls} & {\textbf{Time (s)}} & {\textbf{Tok. (k)}} \\
\midrule
1 & 18 & {100} & 18 & 71(10) & 44(3) & {100} & 18--25 & 196(93) & 98(38) & {100} & 18--24 & 94(46) & 117(49) \\
2 & 18 & {100} & 18 & 70(8) & 44(2) & {100} & 18--24 & 192(60) & 96(24) & {100} & 18--24 & 77(36) & 100(40) \\
3 & 20 & {100} & 20 & 83(4) & 53(0) & {100} & 20--28 & 153(61) & 76(22) & {100} & 20--24 & 72(30) & 102(31) \\
4 & 21 & {100} & 21--32 & 164(53) & 103(34) & {100} & 22--26 & 291(76) & 105(18) & {100} & 21 & 48(3) & 78(5) \\
5 & 41 & {100} & 41 & 141(6) & 137(1) & {100} & 41--70 & 468(377) & 262(150) & {100} & 41--61 & 160(70) & 306(137) \\
6 & 36 & {90} & 32--57 & 231(89) & 163(65) & {100} & 36--47 & 442(131) & 296(83) & {100} & 36--44 & 151(63) & 264(101) \\
7 & 18 & {20} & 258--292 & 1050(106) & 1997(181) & {90} & 35--92 & 421(271) & 355(296) & {100} & 23--77 & 227(77) & 463(192) \\
8 & 18 & {70} & 44--230 & 426(290) & 653(589) & {100} & 25--59 & 344(161) & 190(131) & {100} & 26--52 & 216(143) & 312(163) \\
9 & 21 & {100} & 26--71 & 272(143) & 181(95) & {50} & 35--62 & 673(194) & 523(172) & {40} & 59--88 & 426(101) & 1067(331) \\
\midrule
 &  & {87} & -- & 279(294) & 375(600) & \bfseries 93 & -- & 353(157) & \bfseries 222(143) & \bfseries 93 & -- & \bfseries 163(111) & 312(294) \\
\bottomrule
\end{tabular}
\end{table*}

Table~\ref{tab:results-combined} summarizes the experiment results of the three agent architectures across the nine production challenges, averaged over ten runs per challenge. 
The optimal number of tool calls (Opt) represents a lower bound, assuming the minimum number of tool calls required to fulfill the production order.
However, for Challenges 7, 8, and 9, it should be noted that, while the factory setup theoretically permits the lower bound, it would require prior knowledge of the silent defects.

On the five baseline challenges (1–5), all architectures achieved 100\% SR, confirming that all three designs can handle standard production orders reliably.
All architectures solved Challenge~6 at 90--100\%, demonstrating that concurrently active orders add no fundamental reliability problem to any design.
The more difficult silent-failure Challenges (7–9) expose clear architectural trade-offs.
The orchestrator architecture achieved an overall SR of 87\%.
It performed worst on Challenge~7 (20\% SR), where all magazine blocks of the first module are defective and a fallback to the second module is required.
In the failing runs, the backbone agent entered a loop between the assembly and quality-inspection agents without ever identifying the need to switch modules, consuming an average of 1997\,k tokens and 1050\,s before timing out.
In contrast, the monolithic agent autonomously switched to the alternative module on all ten runs, as confirmed by reasoning traces such as: \textit{``Wait, maybe I should try the other assembly module (edge~2) just in case edge~1 is producing defective blocks?''}, demonstrating emergent fault-diagnosis behavior without explicit fault-handling logic.
However, the orchestrator was the only architecture to solve Challenge~9 at 100\%, as the backbone agent's global view of the transport network allows it to identify the blocked clockwise path and reroute the plate counter-clockwise.

The monolithic and peer-to-peer architectures both achieved the highest overall SR of 93\%.
The monolithic architecture solved all challenges except Challenge~9 (40\%) with the silent conveyor-belt fault between edges~0 and 1.
Because the monolithic policy must jointly handle routing, assembly sequencing, inspection, recycling, and delivery in a single decision loop, these contradictions amplified into long local retry loops and timeouts.
By contrast, the orchestrator benefited from a narrower control problem (routing and coordination), allowing faster strategy correction once the clockwise path proved unreliable.

The peer-to-peer architecture solved both silent assembly module defect Challenges~7 and~8 at 90\% and 100\%, but dropped to 50\% on Challenge~9, where resolving the transport fault requires propagating that information across multiple agents.


In terms of execution time, the monolithic architecture was the fastest overall ($163\pm111$\,s), benefiting from the absence of inter-agent message-passing overhead.
The orchestrator was second ($279\pm294$\,s), with its standard deviation inflated by timeouts on Challenge~7 ($1050\pm106$\,s).
The peer-to-peer architecture was the slowest ($353\pm157$\,s), due to increased message-passing overhead and GPU contention between the concurrent agents.

\begin{figure}[ht]
\centerline{\includegraphics[width=0.5\textwidth]{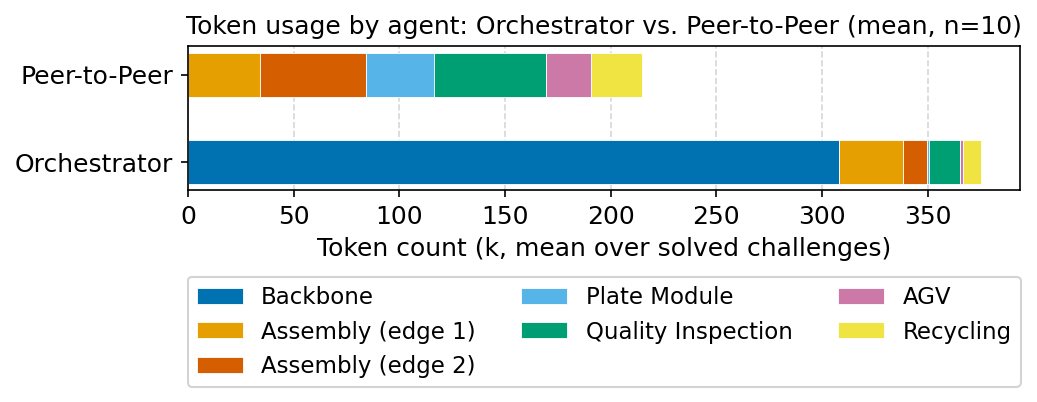}}
\caption{Comparison of mean total token consumption for successful runs across the nine production challenges of the peer-to-peer and orchestrator architecture. Tokens are stacked per agent.}
\label{fig:tokens}
\end{figure}

Token consumption follows a different ordering.
The peer-to-peer architecture consumed the fewest tokens ($222\pm143$\,k), as each specialist agent reasons over a narrow local context.
Despite faster execution, the monolithic architecture with its single expanding context with all tool descriptions and tool outputs consumed $312\pm294$\,k tokens on average, reaching $1067\pm331$\,k on Challenge~9.
The orchestrator consumed the most tokens on average ($375\pm600$\,k), almost entirely driven by Challenge~7, where looping failures produced up to $1997$\,k tokens per run.
On challenges where the orchestrator succeeds reliably (1–6, 9), its token footprint ($44$–$181$\,k) is the lowest of all architectures.
Across all architectures, input tokens accounted for 92–98\% of total token consumption, reflecting the cost of injecting factory state and conversation history into each LLM call.
The comparison of token consumption split by agent in Figure~\ref{fig:tokens} clearly reveals how the peer-to-peer architecture's token consumption is distributed evenly across all specialist agents, while in the orchestrator architecture, the backbone agent dominates token usage, concentrating planning and coordination cost in one place.
For reference, at cloud provider pricing (\$0.12 per 1M input tokens, \$0.35 per 1M output tokens on OpenRouter for Gemma4:31B), costs range from under \$0.01 for simple challenges to approximately \$0.26 for the worst-case looping failure on Challenge 7 ($1997$\,k tokens).

A small ablation study isolates the effect of prompt injecting the current factory state at each inference step (see Table~\ref{tab:monolithic-ablation}).
The ablation is conducted exclusively on the monolithic architecture, as it is the only architecture capable of reconstructing the full factory state from its action history alone.
In the orchestrator and peer-to-peer architectures, each agent's context is inherently partial, making state injection a necessity.
Injecting the factory state improves overall SR from 88\% to 93\%, at the cost of more tokens ($264\,\pm\,203$\,k vs. $312\,\pm\,294$\,k) and slightly longer execution ($128\,\pm\,72$\,s vs. $163\,\pm\,111$\,s).
Without state injection, the LLM must reconstruct the factory state from its action history.
This is feasible for the straightforward challenges, but becomes unreliable when silent defects are present.
The agent cannot distinguish an actual hardware fault from an incorrectly inferred state, causing SR to drop from 40\% to 20\% on Challenge~9 and from 100\% to 90\% on Challenges~7 and~8.

Overall, no single architecture dominates across all metrics.
The monolithic architecture ties the peer-to-peer architecture for the highest mean SR of 93\% and it achieves the fastest execution, making it a strong choice when simplicity and reliability are the primary concerns, though its token cost may limit scalability to larger factory layouts.
The orchestrator architecture excels at belt-routing challenges (100\% on Challenge~9) and achieves the lowest token footprint on challenges it handles well, but its centralized design makes it a single point of failure.
The peer-to-peer architecture offers the most token-efficient operation and competitive SR on isolated-module faults, but it lacks a global replanning mechanism, causing it to underperform when a fault must be diagnosed across agent boundaries, as in Challenge~9.
Extending it with a shared memory bank or a global broadcast channel is a natural direction for future work.

\begin{table}[ht]
\centering
\caption{Ablation: effect of state injection on the monolithic agent (mean over 10 runs, across all challenges). SR = Solve Rate.}
\label{tab:monolithic-ablation}
\begin{tabular}{lccc}
\toprule
\textbf{Variant} & \textbf{SR (\%)} & \textbf{Time (s)} & \textbf{Tokens (k)} \\
\midrule
Monolithic & 93\% & 163$\,\pm\,$111\,s & 312$\,\pm\,$294\,k \\
Monolithic (no injection) & 88\% & 128$\,\pm\,$72\,s & 264$\,\pm\,$203\,k \\
\bottomrule
\end{tabular}
\end{table}

\section{Conclusion}
\label{sec:conclusion}
This paper evaluates the feasibility of a modular smart factory in which production planning and inter-module coordination are driven entirely by LLM-based agents.
We propose an agent architecture in which each factory module is paired with a dedicated LLM agent and an MCP tool server that exposes the module's skills as OPC UA method calls.
Three agent architectures (orchestrator, peer-to-peer, and monolithic) are implemented and evaluated across nine production challenges of increasing complexity, covering multi-module assembly, block rotation, human-in-the-loop, simultaneous multi-plate production, and silent defect detection.

The monolithic architecture and peer-to-peer architecture both achieved the highest mean solve rate of 93\%.
The monolithic architecture succeeded on all challenges except the silent conveyor-belt fault (Challenge~9, 40\%), where the expanding context eventually exhausted the agent's ability to correctly attribute the failure within the tool-call budget.
The peer-to-peer architecture performed well on isolated module-fault scenarios but failed on Challenge~9 (50\%) due to the absence of a global replanning mechanism.
Its fault diagnosis became diluted as it propagated across agent boundaries.
The orchestrator architecture achieved a mean solve rate of 87\%.
It uniquely solved Challenge~9 at 100\% by leveraging its global view of the transport network to reroute plates counter-clockwise around the blocked segment, but failed predominantly on Challenge~7 (20\%), where the backbone agent could not autonomously identify the need to switch assembly modules and entered a costly retry loop.
The ablation study confirmed that real-time factory state injection is critical for reliable fault diagnosis.
Without it the monolithic architecture's solve rate drops from 93\% to 88\%, as the agent struggled to distinguish genuine factory defects from incorrectly inferred state.

The following limitations also outline directions for future work.
First, the prototype was validated exclusively in simulation.
Physical deployment would introduce additional variability, such as sensor noise or communication latency.
Future work should investigate whether smaller, quantized models deployable on industrial edge devices equipped with neural processing units (NPUs) can maintain reliable performance, thereby clarifying the minimum hardware requirements for agent-controlled production.
An alternative to deploying agents on physical hardware is to shift the role of LLM agents from runtime controllers to production program generators.
Rather than issuing tool calls in real time against live hardware, agents would receive a production order, autonomously generate and execute a tool-call sequence within the simulation environment, and validate its correctness against the simulated factory state, as already demonstrated in this work.
Once the generated program has been verified in simulation, it will be presented to a human engineer for review and approval before deployment on physical hardware.

Second, all inter-agent communication runs over a local MQTT broker, meaning the system has not been evaluated under realistic network latency or bandwidth constraints.

Third, the hexagonal conveyor backbone introduces a physical centralization that partially constrains logical decentralization, inherently disadvantaging the peer-to-peer architecture.
More broadly, while all architectures are in principle applicable to other production systems with OPC UA-accessible module skills, scalability to larger factory layouts and more complex module topologies remains an open question.


This is, to our knowledge, the first work to integrate MCP tool servers with OPC UA in a multi-agent LLM framework for factory automation.
The findings suggest that LLM-based agents are a credible candidate for flexible, instruction-driven manufacturing coordination, particularly in settings where reprogramming costs are high and task variability is significant.

\section*{Acknowledgment}

This work is supported by the Horizon Europe Research and Innovation Actions under grant number 101092908 (Smart-Edge).

\bibliographystyle{IEEEtran} 
\bibliography{bibtex} 

@INPROCEEDINGS{Xia,
  author={Xia, Yuchen and Jazdi, Nasser and Zhang, Jize and Shah, Chaitanya and Weyrich, Michael},
  booktitle={2025 IEEE 30th International Conference on Emerging Technologies and Factory Automation (ETFA)}, 
  title={{Control Industrial Automation System with Large Language Model Agents}}, 
  year={2025},
  volume={},
  number={},
  pages={1-8},
  doi={10.1109/ETFA65518.2025.11205539}}

@INPROCEEDINGS{MCPinManufacturing,
  author={Vieira Da Silva, Luis Miguel and Köcher, Aljosha and Gehlhoff, Felix},
  booktitle={2025 IEEE 30th International Conference on Emerging Technologies and Factory Automation (ETFA)}, 
  title={{Beyond Formal Semantics for Capabilities and Skills: Model Context Protocol in Manufacturing}}, 
  year={2025},
  volume={},
  number={},
  pages={1-4},
  doi={10.1109/ETFA65518.2025.11205601}}

@online{supcon2025factoryagent,
  author    = {SUPCON Global},
  title     = {{Building a Decision-Making LLM Agent for a Simulated Factory}},
  year      = {2025},
  month     = oct,
  url       = {https://global.supcon.com/posts/building-a-decision-making-llm-agent-for-a-simulated-factory-1},
}

@online{supcon2025mqtt,
  author    = {SUPCON Global},
  title     = {{Why {MQTT} Topics + {JSON} Are Ideal for {LLM} Agents in a Simulated Factory}},
  year      = {2025},
  month     = sep,
  url       = {https://global.supcon.com/posts/why-mqtt-topics-json-are-ideal-for-llm-agents-in-a-simulated-factory},
  note      = {Accessed: 2026-02-23}
}

@misc{LLMandOPCUA,
      title={{Beyond touch-based human-machine interface: Control your machines in natural language by utilizing large language models and {OPC UA}}}, 
      author={Bernd Hofmann and Niklas Piechulek and Sven Kreitlein and Joerg Franke and Patrick Bruendl},
      year={2026},
      eprint={2510.11300},
      archivePrefix={arXiv},
      primaryClass={cs.HC},
      url={https://arxiv.org/abs/2510.11300}, 
}

@INPROCEEDINGS{LLMasController,
  author={Xu, Jiahong and Zheng, Zhiwei and Wang, Zaijun},
  booktitle={2024 IEEE International Conference on Robotics and Biomimetics (ROBIO)}, 
  title={{{LAC}: Using LLM-based Agents as the Controller to Realize Embodied Robot}}, 
  year={2024},
  volume={},
  number={},
  pages={1894-1899},
  doi={10.1109/ROBIO64047.2024.10907494}}

@INPROCEEDINGS{VirtuallyCommissionedSemanticRecipes,
  author={Köhle, Kay and Anicic, Darko and Runkler, Thomas and Dorofeev, Kirill and Thuluva, Aparna and Graf, René},
  booktitle={2025 IEEE 8th International Conference on Industrial Cyber-Physical Systems (ICPS)}, 
  title={{Virtually Commissioned Semantic Recipes based on OPC UA Skills}}, 
  year={2025},
  volume={},
  number={},
  pages={1-6},
  doi={10.1109/ICPS65515.2025.11087832}}

@INPROCEEDINGS{skills,
  author={Zimmermann, Patrick and Axmann, Etienne and Brandenbourger, Benjamin and Dorofeev, Kirill and Mankowski, André and Zanini, Paulo},
  booktitle={2019 24th IEEE International Conference on Emerging Technologies and Factory Automation (ETFA)}, 
  title={{Skill-based Engineering and Control on Field-Device-Level with OPC UA}}, 
  year={2019},
  volume={},
  number={},
  pages={1101-1108},
  doi={10.1109/ETFA.2019.8869473}}

@ARTICLE{IoTProtocolSurvey,
  author={Al-Fuqaha, Ala and Guizani, Mohsen and Mohammadi, Mehdi and Aledhari, Mohammed and Ayyash, Moussa},
  journal={IEEE Communications Surveys \& Tutorials}, 
  title={{Internet of Things: A Survey on Enabling Technologies, Protocols, and Applications}}, 
  year={2015},
  volume={17},
  number={4},
  pages={2347-2376},
  doi={10.1109/COMST.2015.2444095}}

@article{Pulikottil2023AgentManufacturing,
  author    = {Pulikottil, Terrin and Estrada-Jimenez, Luis A. and Ur Rehman, Hamood and Mo, Fan and Nikghadam-Hojjati, Sanaz and Barata, Jose},
  title     = {{Agent-based manufacturing---review and expert evaluation}},
  journal   = {The International Journal of Advanced Manufacturing Technology},
  volume    = {127},
  number    = {},
  pages     = {2151--2180},
  year      = {2023},
  doi       = {10.1007/s00170-023-11517-8},
  publisher = {Springer},
}

@INPROCEEDINGS{Lim2024,
  author={Lim, Jonghan and Vogel-Heuser, Birgit and Kovalenko, Ilya},
  booktitle={2024 IEEE 20th International Conference on Automation Science and Engineering (CASE)}, 
  title={{Large Language Model-Enabled Multi-Agent Manufacturing Systems}}, 
  year={2024},
  volume={},
  number={},
  pages={3940-3946},
  doi={10.1109/CASE59546.2024.10711432}}

@article{Zhao2026,
title = {{A Large language model-based multi-agent manufacturing system for intelligent shopfloors}},
journal = {Advanced Engineering Informatics},
volume = {69},
pages = {103888},
year = {2026},
issn = {1474-0346},
doi = {https://doi.org/10.1016/j.aei.2025.103888},
url = {https://www.sciencedirect.com/science/article/pii/S1474034625007815},
author = {Zhen Zhao and Dunbing Tang and Changchun Liu and Liping Wang and Zequn Zhang and Haihua Zhu and Kai Chen and Qingwei Nie and Yuchen Ji},
}

@misc{yao2023react,
      title={{ReAct: Synergizing Reasoning and Acting in Language Models}}, 
      author={Shunyu Yao and Jeffrey Zhao and Dian Yu and Nan Du and Izhak Shafran and Karthik Narasimhan and Yuan Cao},
      year={2023},
      eprint={2210.03629},
      archivePrefix={arXiv},
      primaryClass={cs.CL},
      url={https://arxiv.org/abs/2210.03629}, 
}

@misc{opcua,  
    author = {{OPC Foundation}},  
    title = {{OPC UA Specification}},  
    year = {2024},  
    publisher = {Open Platform Communications (OPC) Foundation},  
    url = {https://opcfoundation.org/},  
    note = {Accessed: Nov. 26, 2024}  
}

@misc{mcp2024,
  author       = {Anthropic},
  title        = {{Model Context Protocol Specification}},
  year         = {2024},
  howpublished = {\url{https://spec.modelcontextprotocol.io}},
  note         = {Accessed: 2025}
}


\end{document}